\documentclass[%
 reprint,
 amsmath,amssymb,
 aps,
]{revtex4-1}

\def\MET{$\not$E$_t$~}
\usepackage{graphicx}
\usepackage{dcolumn}
\usepackage{bm}

\begin{document}


\title{ATLAS Overview and Main Results}%

\author{Krzysztof Sliwa}
\affiliation{Tufts University, Medford, Massachusetts 02155, USA}
\collaboration{On behalf of the ATLAS Collaboration}


\date{May 13, 2013}

\begin{abstract}
An overview of the ATLAS experiment, its physics program and a selection of the most important results, based on the data taken in pp collisions at energies of 7 and 8 TeV in 2011 and 2012, respectively, is presented. The question of possible changes in our understanding of elementary particles physics, after a discovery of a new boson of the mass of $\sim 125$ GeV last summer, is addressed. During the current long shutdown, the Large Hadron Collider (LHC) will be upgraded to allow the LHC experiments to study pp collisions at the energy of $\sim 13~$TeV.The ATLAS plans for future analyses and measurements with the new data to be taken after 2015, are summarized. 
\end{abstract}

\maketitle


\section{\label{sec:level1} Physics program at the LHC and ATLAS}

The two most important objectives of the physics program of the Large Hadron Collider at CERN are: i) to perform as many as possible precise measurements, compare them to the Standard Model predictions and, perhaps, find deviations which could provide evidence for "physics beyond the Standard Model"; ii) to find experimental clues to the mechanism of spontaneous breaking of the electroweak symmetry, which remains the only untested part of the Standard Model -  to find Higgs, or multiple Higgs (if they exist), and to measure their properties - or to exclude their existence.

The Standard Model (SM) of particle physics\cite{Glashow}, developed in circa 1975, is a gauge theory based on the $SU(3)_c \times SU(2)_I \times U(1)_Y$ "internal" symmetries. ("c" stands for color, "I"  for weak isospin and "Y" for hypercharge). The $SU(3)_c$ is an unbroken symmetry; it gives rise to Quantum Chromodynamics (QCD), a quantum theory of strong interactions, whose carriers (gluons) are massless and couple to colour - the strong force charge. The $SU(2)\times U(1)$ symmetries, which yield a quantum theory of electroweak interactions, are spontaneously broken, according to what is frequently called the Brout-Englert-Higgs mechanism\cite{Higgs}, which gives masses to the electroweak bosons (massive $W^+,W^-,Z^o$ and the massless photon) and all fermions. In the Minimal Standard Model (MSM), the Higgs sector is the simplest possible. It contains a single weak isospin doublet of complex Higgs fields which, after giving masses to $W^+,W^-,Z^o$, leaves a single neutral Higgs particle which should be observed. Matter is made of fermions - three families (weak isospin doublets) of quarks and three families of leptons, all with corresponding antiparticles. Quarks exist in three colours, while  leptons are colour singlets -  they don't couple to gluons. Bosons are carriers of interactions: there are 8 massless gluons, 3 heavy weak bosons $(W^+,W^-,Z^o)$, and 1 massless photon. A neutral scalar Higgs field permeates the Universe and is, somehow, responsible for masses of all other particles - they originate from couplings to the Higgs field.
In the Minimal Standard Model, this neutral Higgs scalar is the only missing particle.

On July 4th, 2012, the ATLAS and CMS Collaborations announced the discovery of a new particle\cite{discovery}, a Higgs-like boson with the mass of of $\sim 125$ GeV. It was a great day for CERN, ATLAS and CMS physicists, and a great result for the more than 20 years old LHC project. Fantastic result that it is, it actually brings immediately many questions, some new, some old: i) is this the Minimal Standard Model Higgs? Answering this question will take time and many precision measurements - with the Higgs mass known, all MSM couplings can now be calculated and compared with the experimental results; ii) there remain {\it many} unsolved problems in SM. There is still plenty to understand and to search for, even if the new discovered boson is the MSM Higgs. 

Personally, I think it would be more interesting if the elementary Higgs boson were not there, or if the new-found particle is {\it not} a MSM Higgs boson. 
\section{\label{sec:level1} Standard Model - outstanding questions}

The SM leaves many questions unanswered: i) why are there so many (26) free parameters - all masses, couplings, mixing angles and CP-violating phases - all have to be measured; ii) why are there 6 quark and 6 leptons - is there an additional symmetry? iii) why is CP not an exact symmetry (or why are laws of physics not symmetric between matter and antimatter? The problem is, perhaps, related to the question why is our Universe matter-dominated, however, there does not seem to be enough CP-violation in SM -  what is its origin? iv) SM does not provide any clues about "dark matter", which seems to be about 5 times more prevalent (27\% vs 5\%) in the Universe than ordinary matter\cite{Planck}; v) how to include gravity? It is only natural to think that Standard Model is just a low-energy effective theory, an approximation.

Another difficulty is related to spontaneous symmetry breaking - the heart of the SM. The problem appears if the Higgs field is an elementary scalar. The quantum corrections to scalar particle (Higgs) mass exhibit quadratic dependence of the cutoff scale $\Lambda$, making the Higgs mass {\it very} sensitive to the scale of {\it new} physics. This is known as {\it fine tuning problem} or a {\it gauge hierarchy problem}; as the fine-tuning of parameters has to be performed order by order in perturbation theory - a very unpleasant feature of the MSM. The original problem - how to give masses to weak bosons in a gauge invariant way - was only partially solved by the Higgs mechanism, and the problem was transferred to a new level, where the new puzzle is how to keep Higgs mass stable against large quantum corrections from the higher energy scales. A method of controlling Higgs mass divergencies other than fine tuning of parameters would be very welcome.
\subsection{\label{sec:level2}Supersymmetry - the most elegant solution? }

Supersymmetry is a space-time symmetry which introduces a fermionic partner to every boson and vice-versa, identical in all quantum numbers other than spin. The Higgs mass divergencies would cancel, without any fine tuning, in all orders of perturbation theory as the quantum loop corrections from bosons and fermions come with opposite sign.
If Supersymmetry (SUSY) were real, it must be somehow broken, as we have not yet observed super particles, while still keeping the ability to solve the gauge hierarchy problem. This is not an easy task, it depends on the scale at which SUSY is broken, and on how it is broken. To some extent, it remains an open question. SUSY provides a natural explanation for "dark matter". Local supersymmetry could also provide a viable theory of gravity - supergravity.
\subsection{\label{sec:level2}Gauge theories and extra dimensions }

Gauge theories can be understood best in the mathematical language of fibre bundles\cite{Atiyah} - a gauge potential (e.g. 4-vector potential of electrodynamics, or Yang-Mills potential for electroweak theory) is a connection in the principal fibre bundle, a state-space described by a given gauge group (U(1) of electrodynamics, SU(2) of Yang-Mills theory), superimposed on space-time. The curvature of the connection is the gauge field, for example, the field-strength tensor $F_{\mu \nu}$ of electrodynamics. In such a geometrical picture, the strong and electroweak interactions are very similar to Einstein's gravity, except the distortion measured by curvature is not taking place in the geometry of space-time, but in the geometry of the higher-dimensional "total-space", imposed over space-time. Gauge (or, rather, phase transformations) are analogous to co-ordinate transformations in Riemannian geometry of Einstein's General Relativity. We may be living in a world which is more than four-dimensional (10, perhaps 11?), except that we don't "see" beyond the familiar four space-time dimensions\cite{Scherk}. 
\subsection{\label{sec:level2}Beyond Standard Model?}

The list of possibilities is quite long: 
\begin{itemize}
\item Supersymmetry
\item Grand Unified Theories based on larger symmetry groups (e.g. $SU(5), SU(10), E_8\times E_8,$ Monster Group)
\item new models - extensions of Kaluza-Klein theory, string theory, superstring theories, branes, M-theory, loop quantum gravity...
\item Technicolor, other models of dynamical symmetry breaking?
\end{itemize}
Finding the Higgs is a very important physics result, however, it does {\it not} solve the SM problems and questions. To try to answer them, more experimental data is badly needed.
\section{\label{sec:level1} Large Hadron Collider at CERN}

The LHC is a superconducting proton accelerator and collider installed in a 27 km circumference underground tunnel at CERN. The tunnel (cross section diameter 4 m) was built for LEP collider in 1985. The first studies for a high-energy proton-proton collider in the LEP tunnel started already in 1984. LEP2 was closed in 2000, and installation of LHC machine and experiments started in 2003. The first collisions at $\sqrt(s)=900$ GeV were recorded on November 23, 2009. The first collisions at $\sqrt(s)=7~$TeV took place on March 30, 2010, starting a long physics program. First hints of a new particle of mass ~$125$ GeV were reported by the end of 2010. First collisions at $\sqrt(s)=8~$TeV were recorded on May 1st, 2012, and the machine was performing spectacularly, delivering fast increasing luminosities. On July 4th, 2012 a discovery of a Higgs-like boson was announced. 
\section{\label{sec:level1} ATLAS DETECTOR AT LHC and ATLAS Collaboration}

\begin{figure}[h]
\includegraphics[width=0.5\textwidth]{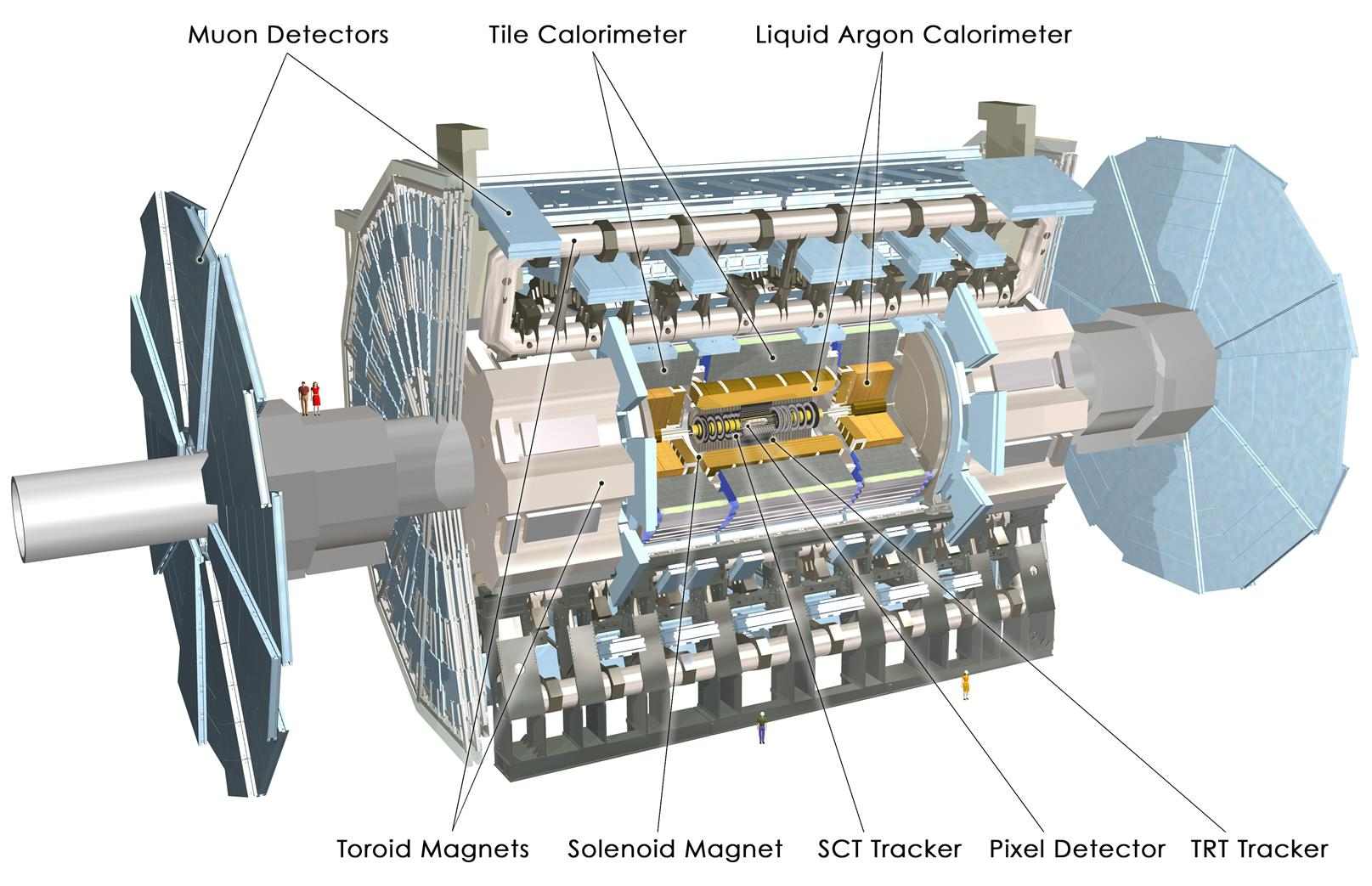}
\caption{\label{fig:ATLAS} A schematic drawing of the ATLAS detector.}
\end{figure}

ATLAS is a more than 20 years project, its Letter of Intent was prepared in 1992. It is a large, multipurpose particle detector\cite{ATLAS} located at the LHC Point 1, very close to CERN Meyrin site. It has been designed to measure, reconstruct and identify leptons, photons, quark jets, individual charged particles (tracks) and primary and secondary vertices. ATLAS detector is 44 m  wide, 22 m high, weighs 7000 t, and its layout is shown in Fig.\ref{fig:ATLAS}. The main components of the ATLAS detector is briefly described providing, in this way, an introduction to many other ATLAS talks at this conference.

\begin{itemize}

\item The ATLAS detector is equipped with 4 superconducting magnets: a solenoid, located inside of the calorimeters, which provides a 2 T magnetic field for the tracking detectors, and three toroids - an air-core central and two end-cap toroids, which provide magnetic fields of 0.5-1.5 T in a large volume outside of the calorimeters to detect, identify and measure muons. 

\item The ATLAS Inner Detector is located inside a solenoid which provide 2 T magnetic field. The detector elements closest to the beam are only about 5 centimeters from the beam axis. The detector itself has a radius of 1.2 m, and is 6.2 m long. Moving from the innermost to the outermost of the detector, the barrel tracker consists of silicon pixel detectors, silicon strip detectors  (SCT) and a transition radiation tracker (TRT). The end-cap system consists of transition radiation detectors and silicon strips. The ATLAS Inner Detector provides precise tracking and vertexing, particle identification and $e/\pi$ separation. Its momentum resolution is $\sigma /p_T \sim 3.8\times 10^{-4}~ p_T$(GeV)$ \oplus 0.015$.

\item
The ATLAS calorimeters, shown in Fig.\ref{fig:CALO}, are located outside of  the solenoid. The ATLAS electromagnetic calorimeter is based on Pb-LAr (lead-liquid argon) technology, with accordion layout. It provides trigger, $e/\gamma$ identification and measurement with energy resolution of $\sigma/E \sim 10\%/ \sqrt(E)$ and angular coverage in pseudorapidity range $|\eta|<3.2$. The forward liquid argon calorimeter provides coverage in the range $3.1 < |\eta| < 4.9$ and the energy resolution of $\sigma/E \sim 29\%/ \sqrt(E) \oplus 0.04.$ The hadronic central calorimeter is based on Fe/scintillator tiles. It provides coverage up to $|\eta|<1.7$ and the energy resolution of  $\sigma/E \sim 50\%/\sqrt(E) \oplus 0.03.$ The end-cap calorimeters use Cu/W-LAr design, with the angular coverage of $1.5 < |\eta|<3.2$ and the energy resolution of $\sigma/E \sim 95\%/\sqrt(E) \oplus 0.08.$ The total mass of the ATLAS calorimeters is about 4000 t.

\begin{figure}[h]
\hspace{1.0cm}
\includegraphics[width=0.5\textwidth]{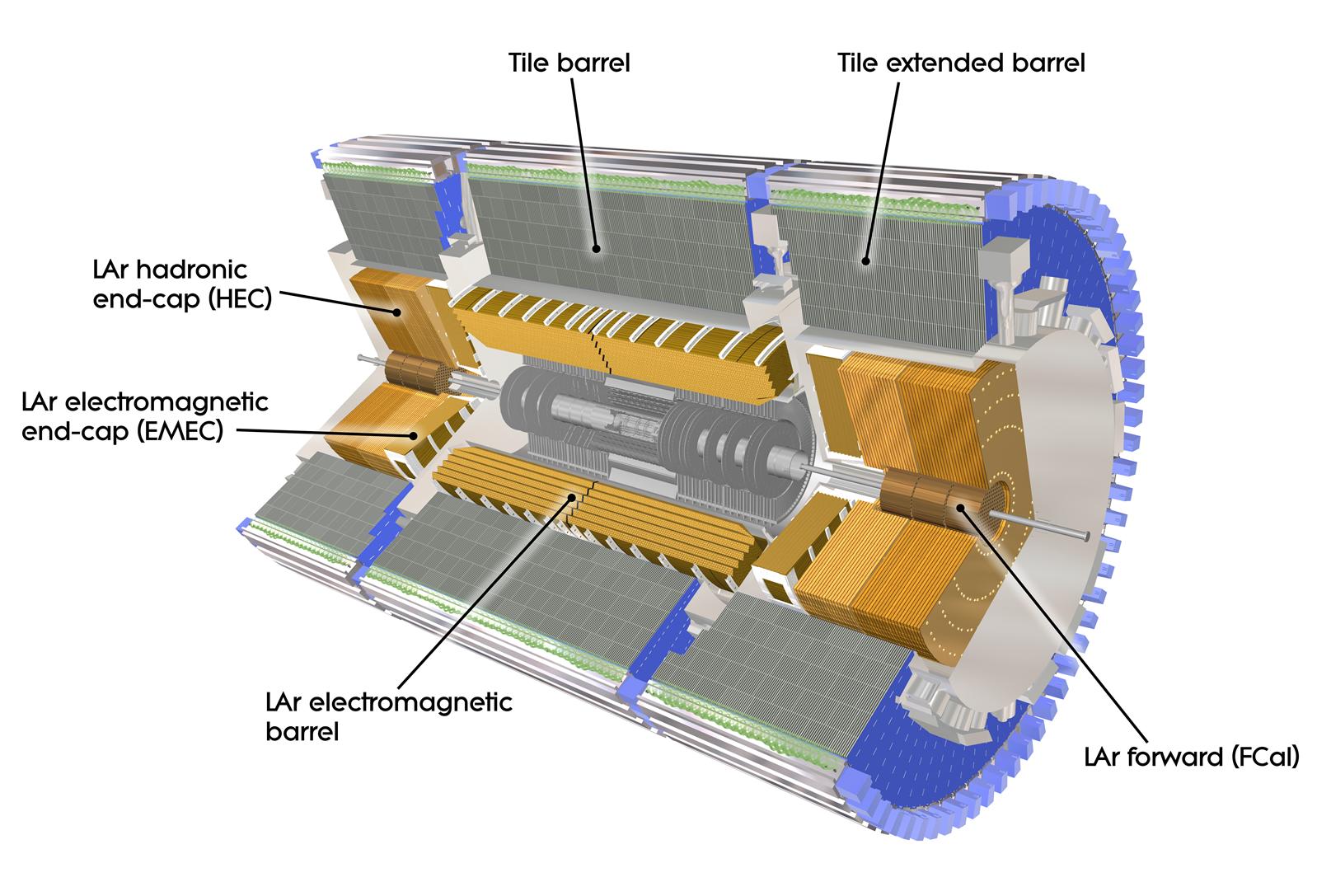}
\caption{\label{fig:CALO} A cut-out view of ATLAS calorimeter detectors.}
\end{figure}

\item
The ATLAS standalone muon spectrometer, shown in Fig.\ref{fig:MUONS},  consists of gas-based muon chambers positioned within a large volume magnetic field created by 3 air-core toroids. It provides muon momentum measurement within $|\eta|<2.7$. Muons are measured with tracking chambers based on CSC (cathode strip) and MDT (monitored drift tubes) technologies, with momentum resolution $<10\%$ up to $E_\mu \sim 1 $ TeV. Trigger information is provided by TGC (thin-gap) and RPC (resistive-plate) chambers. 

\begin{figure}[h]
\hspace{1.0cm}
\includegraphics[width=0.47\textwidth]{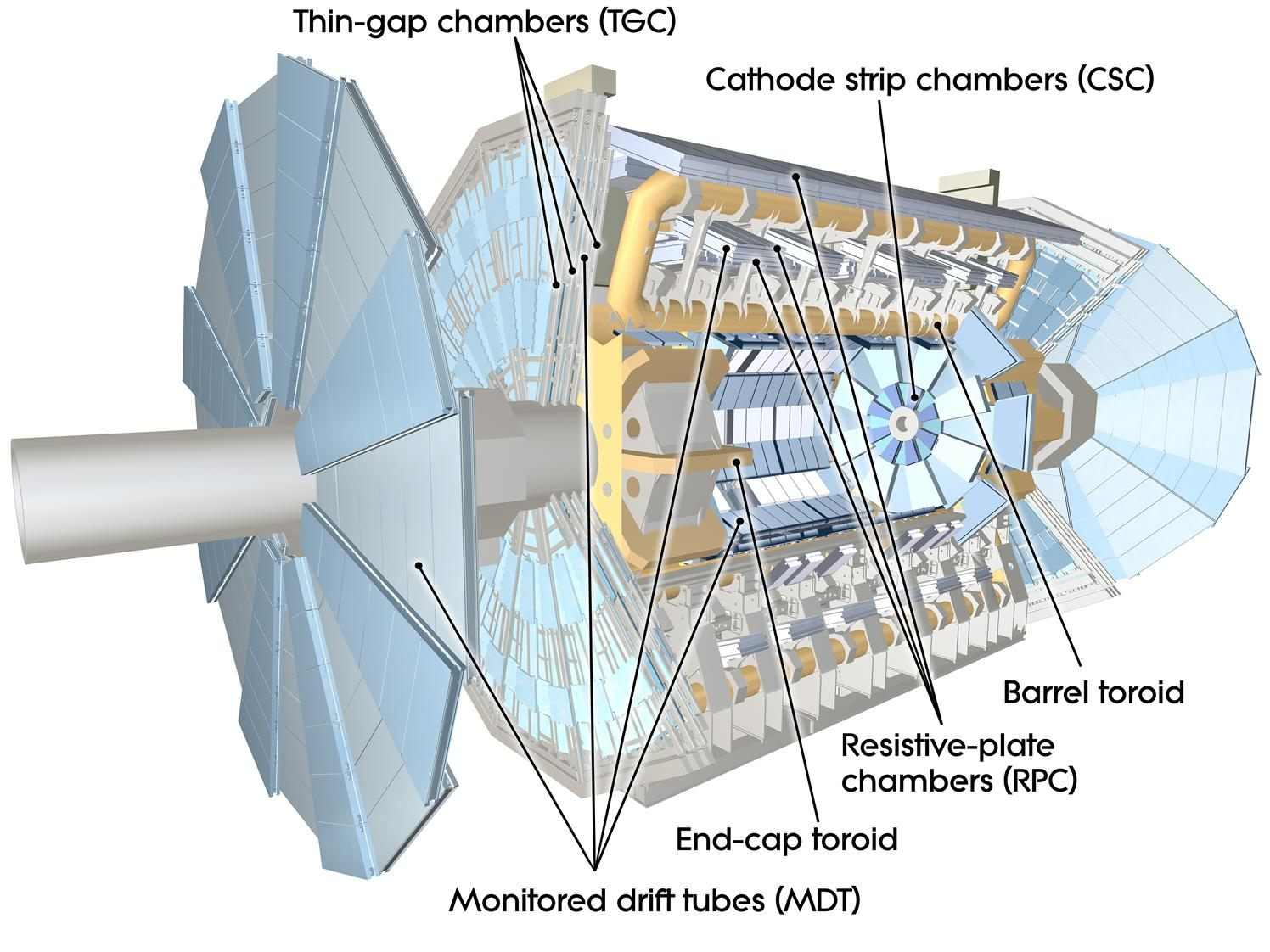}
\caption{\label{fig:MUONS} ATLAS Muon Spectrometer magnets and detectors.}
\end{figure}

\item The ATLAS Trigger and acquisition system faces the formidable task of  selecting only the interesting events from the initial interaction rate of $\sim 400~MHz$ to $\sim 200-400 $ Hz rate with which events are stored and sent to offline processing and reconstruction. The three stage selection system consists of: i) Level1 Trigger (fast on-line electronics, with the decision time $<2.5~\mu s$) and High Level Trigger based on large computer farms: ii)  Level2 Trriger with average decision time $\sim 40~ ms$ per events; and iii) the Event Filter, which processes complete event buffers it receives from the Event Builder, based on 1800 computer nodes, with an average decision time $\sim 4~s$ per event. The number of computing cores used in 2012 run was 17000, the peak event building bandwidth  $10~GB/s$, the peak storage bandwidth $16~GB/s$ and the amount of data recorded reached $6~PB$.

\item The ATLAS computing is a distributed hierarchical system, with Tier-0 at CERN where the data is recorded into tape, and the first pass reconstruction is  performed; 10 Tier-1 centers which participate in reprocessing and Monte Carlo simulation production and about 70 Tier-2 centers, which participate in MC production and support user analyses. The system uses world-wide computer grid systems; most of user analyses are performed off-grid in local Tier-3 computer centers.

\item The ATLAS Collaboration author list contains $\sim 3000$ names of physicists (including the names of about 1000 students). These physicists from 48 countries and 177 universities, working on understanding the detector, calibrations and on physics analyses. There are also hundreds of engineers, technicians and computer scientists whose names don't appear on ATLAS papers. ATLAS results are a truly collaborative effort of many thousands of people supported by many funding agencies from all over the world.
\end{itemize}
\section{\label{sec:level1} FIRST ATLAS PHYSICS RUN 2009-2012}

The LHC performance in its first physics run exceeded expectations, it was fantastic. 

\begin{figure}[h]
\hspace{0.4cm}
\includegraphics[width=0.47\textwidth]{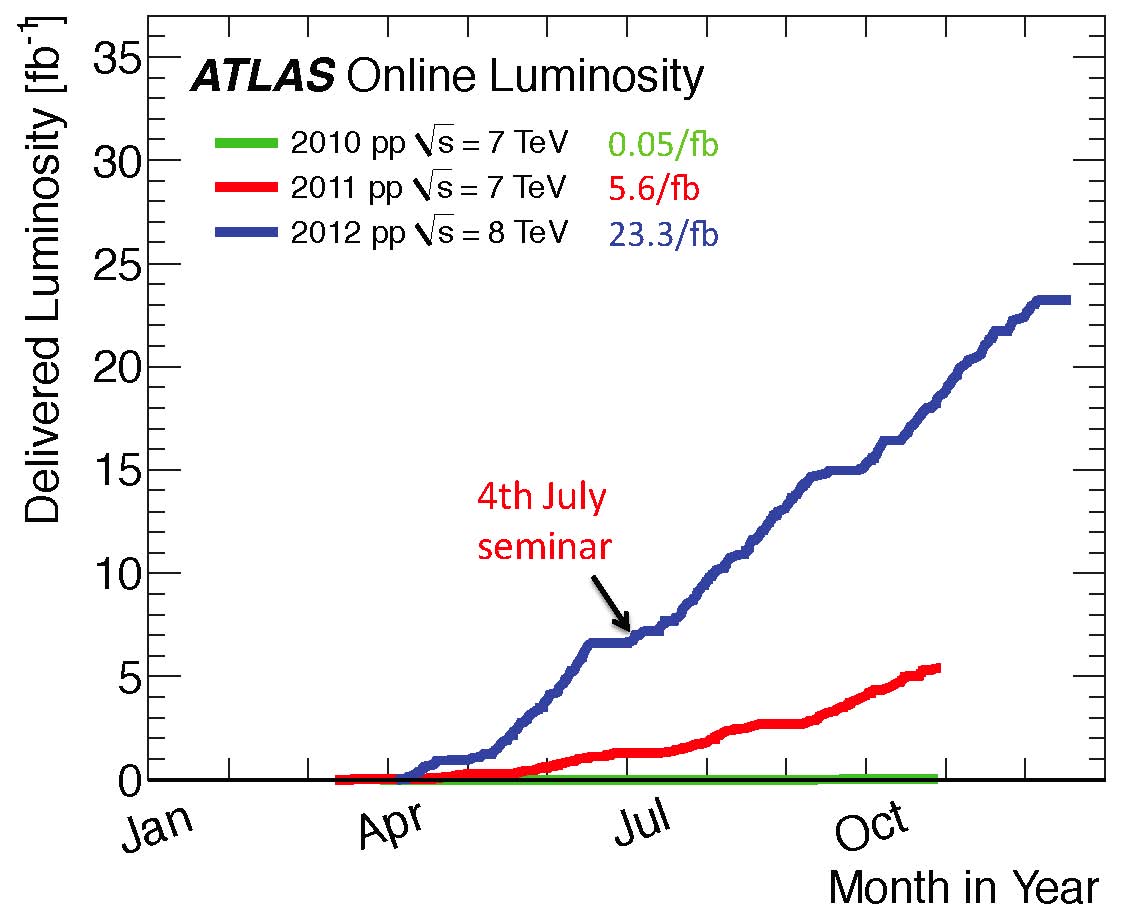}
\caption{\label{fig:LUMI} ATLAS delivered luminosity in 2010 (green), 2011 (red) and 2012 (blue) as a function of month in the year. The total delivered luminosity was almost 29/fb.}
\end{figure}

\begin{figure}[h]
\hspace{1.0cm}
\includegraphics[width=0.49\textwidth]{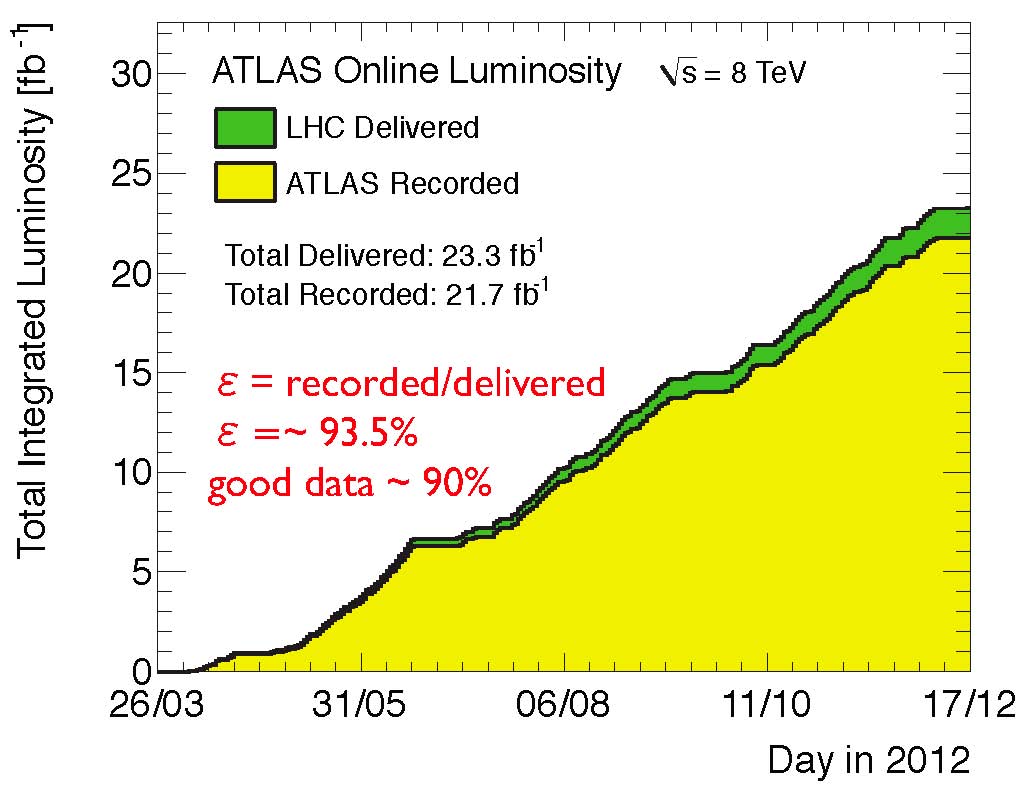}
\caption{\label{fig:EFF} ATLAS recorded luminosity as a function of time in 2012 physics run at $8~$TeV.}
\end{figure}

\begin{figure}[h]
\hspace{1.0cm}
\includegraphics[width=0.49\textwidth]{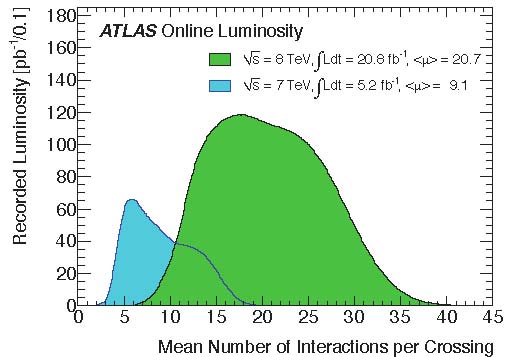}
\caption{\label{fig:MU} Average number of interactions per beam crossing, a measure of pile-up.}
\end{figure}

The delivered luminosity as a function of time for the data taking periods in 2010 and 2011 at $7~$TeV, and in 2012 at $8~$TeV is shown in Fig.\ref{fig:LUMI}. The maximum peak luminosity in 2012 running reached $7.7 \times 10^{33} s^{-1}cm^{-2}$ in August. In July, the LHC recorded its longest stable beams period, which lasted 22.8 h, and the record weekly data-taking efficiency of 55\% was set, as well. 
The detector operation efficiency and data quality in 2012 running are summarized in Fig.\ref{fig:EFF}. 

To achieve such high luminosities, the LHC was operating with 50 ns proton bunch spacing, rather than the nominal 25 ns. This resulted in twice bigger pile-up for the same luminosity, and the peak number of interactions per beam crossing , $<\mu>$, significantly exceeded the design value for $L=10^{34} s^{-1}cm^{-2}$. The distribution of the average number of interactions per beam crossing, $<\mu>$, is presented in Fig.\ref{fig:MU}. It is quite clear that the pile-up problem was much more serious in the $8~$TeV run in 2012 than in the $7~$TeV run in 2011, with $<\mu>$ rising from $\sim 9$ to $\sim 21$. A lot of effort was devoted to prepare 2012 operations. Trigger and off-line algorithms which are pile-up "robust" needed to be developed. In general, pile-up has sizable impact on trigger rates and also on jets, missing transverse energy - \MET and tau lepton reconstruction. Higher trigger rates and larger event size due to pile-up resulted in the increase in the reconstruction time - from 10 s/event for $<\mu>=5$ to 50 s/event for $<\mu>=50$, which represented a challenge for ATLAS offline computing. Also, significant improvements were made in MC modeling of the in-time and out-of-time pile-up effects. No significant pile-up effect on tracking, reconstruction of muon, electrons and photons was found. 

\begin{figure}[h]
\hspace{0.4cm}
\includegraphics[width=0.48\textwidth]{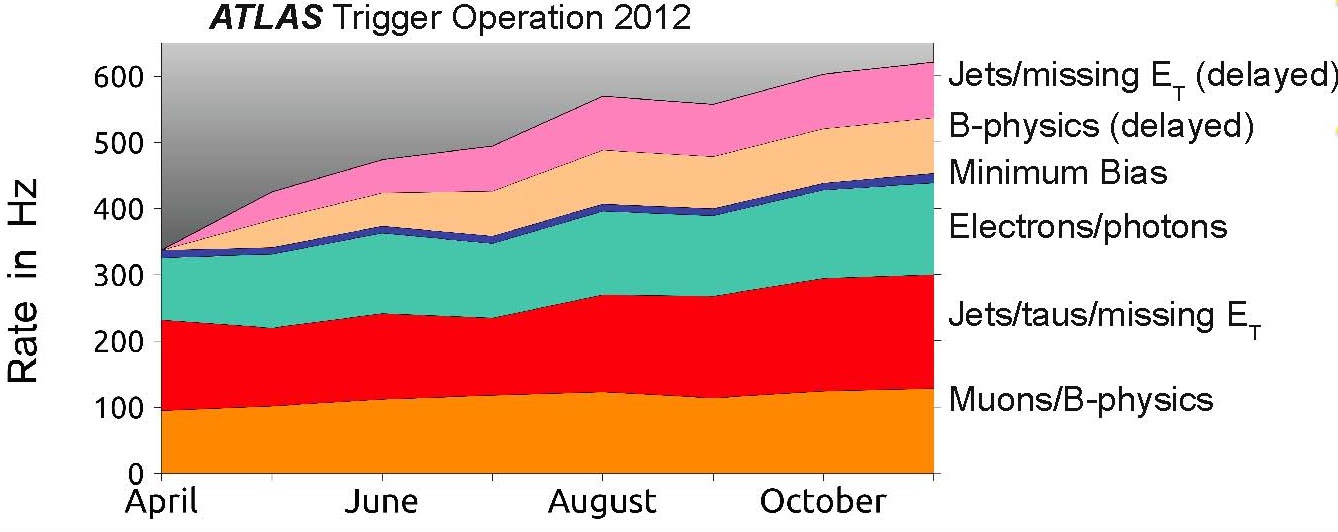}
\caption{\label{fig:triggers} The composition of prompt trigger streams and delayed trigger streams in 2012 physics run.}
\end{figure}

The baseline trigger menu has been designed for  $L=8\times10^{33} s^{-1}cm^{-2}$, and remained mostly unchanged through the 2012  run at $8~$TeV. The contribution of different physics streams to the out-of-Event Filter rate is shown in Fig.\ref{fig:triggers}. The average rate of events which were econstructed promptly at Tier-0 at CERN was $\sim$400 Hz, while reconstruction of the Jets/\MET and B-physics triggers was delayed. 

More details, 
plots demonstrating ATLAS's understanding of pile-up effects on the Level1, Level2 and Event Filter trigger efficiencies, 
can be found in the complete set of slides of the talk\cite{slides}.
\section{\label{sec:level1} ATLAS - MAIN RESULTS}

In this paper I am going to present only a selection of results from a large set of ATLAS analyses. For more details about other ATLAS results please consult other ATLAS talks at this conference by Nicola Orlando\cite{nicola}, Marilyn Marx\cite{marilyn}, Carsten Hensel\cite{carsten}, Elisa Pueschel\cite{elisa}, Sofia Maria Consonni\cite{sofia}, Edson Carquin\cite{edson} and Marisilvia Donadelli\cite{marisilvia}. The ATLAS physics results can be divided into three categories:

\begin{itemize}
\item
Precision measurements and tests of Standard Model - QCD studies; WW, WZ, ZZ, W$\gamma$, Z$\gamma$, $\gamma \gamma$, t$\bar{t}$, single top - these processes are interesting on their own, but also constitute the most important backgrounds to most Higgs and new physics searches. These measurements are well underway.

\item  

Searches for physics "beyond the Standard Model" - so far, no indication of any new physics (other than a new Higgs-like boson) has been found

\item
Higgs searches - a new boson has been found at $\sim 125$ GeV in 2012. This important discovery, however, brings immediately many questions: is it just one Higgs? more than one? two? what is its spin-parity? is this the MSM Higgs boson? 
\end{itemize}
\subsection{\label{sec:level2}QCD jet studies} 

\begin{figure}[h]
\includegraphics[width=0.47\textwidth]{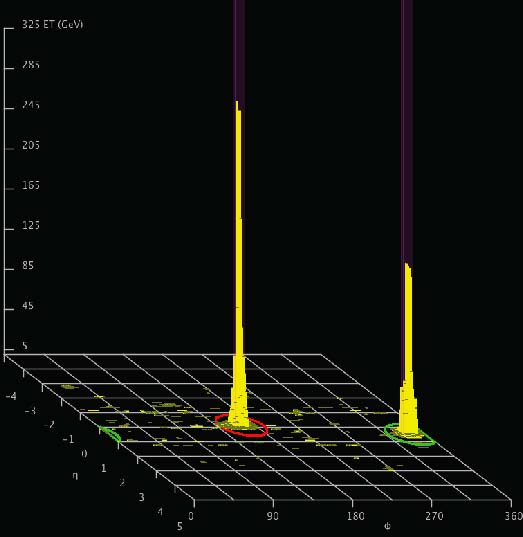}
\caption{\label{fig:JETS} The calorimeter display of an event recorded August 31, 2012. The two central high-$p_T$ jets have the invariant mass of M(jj)=4.69 TeV, the highest recorded ever.}
\end{figure}

\begin{figure}[h]
\includegraphics[width=0.45\textwidth]{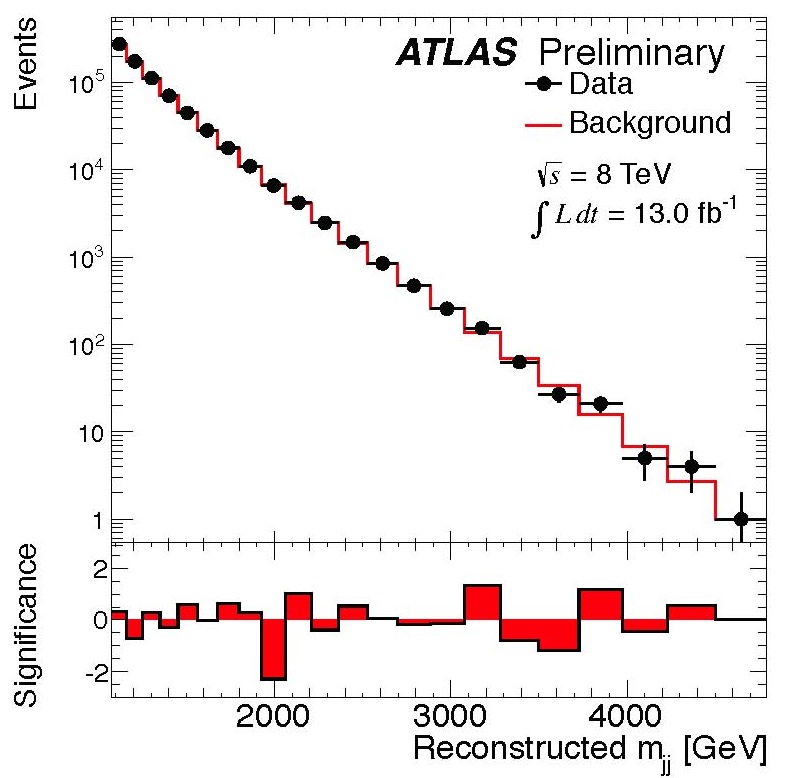}
\caption{\label{fig:JJMass} The M(jj) spectrum based on data from 2012 run at  $8~$TeV, together with the background estimate\cite{mq*}.}
\end{figure}

Understanding of QCD effects is essential for all precision measurements, as QCD backgrounds dominate most analyses. QCD studies cover wide range of phonemena. I chose examples of: i)  analyses of events with highest transverse momentum, $p_T$, jet systems - di-jets and mono jets; ii)  studies of individual, very low $p_T$ tracks.  A two-dimensional plot, pseudorapidity $\eta$ versus azimuth angle $\phi$, of the energy deposited in ATLAS calorimeter towers for  two central high-$p_T$ jets with an invariant mass of 4.69 TeV, the highest recorded in 2012 running at $8~$TeV, is shown in Fig.\ref{fig:JETS}. The distribution of the mass of two-jet system, M(jj), is presented in Fig.\ref{fig:JJMass}. The lack of excess of events at high M(jj) can be translated into a limit.\cite{mq*} on the mass of excited quarks - M(q$^*)< 3.84~$TeV at 95\% confidence level (C.L.). 

\begin{figure}[h]
\includegraphics[width=0.47\textwidth]{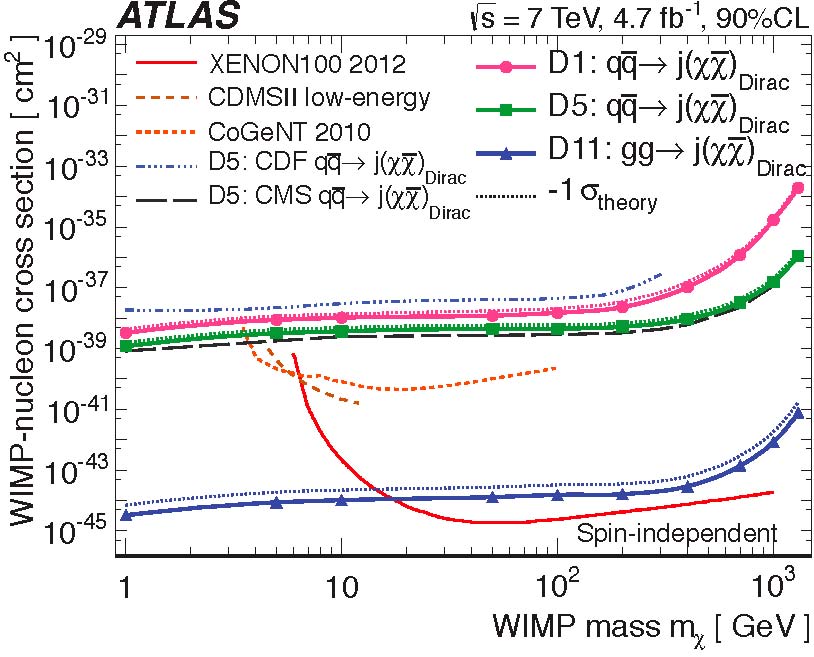}
\caption{\label{fig:WIMP} Limits on the interaction cross section between a dark matter candidate - weak interacting massve particle (WIMP) - and a nucleon, based on an interpretation of a monojet analysis\cite{WIMP}.}
\end{figure}

A relatively simple analysis of events with a single, high-$p_T$, jet production has been interpreted\cite{WIMP} in terms of WIMP-nucleon cross section as a function of the WIMP mass. The ATLAS limits are shown in Fig.\ref{fig:WIMP}. For masses larger than a few GeV, the ATLAS spin-independent limits are more stringent than those obtained in laboratory experiments.

\begin{figure}[h]
\includegraphics[width=0.47\textwidth]{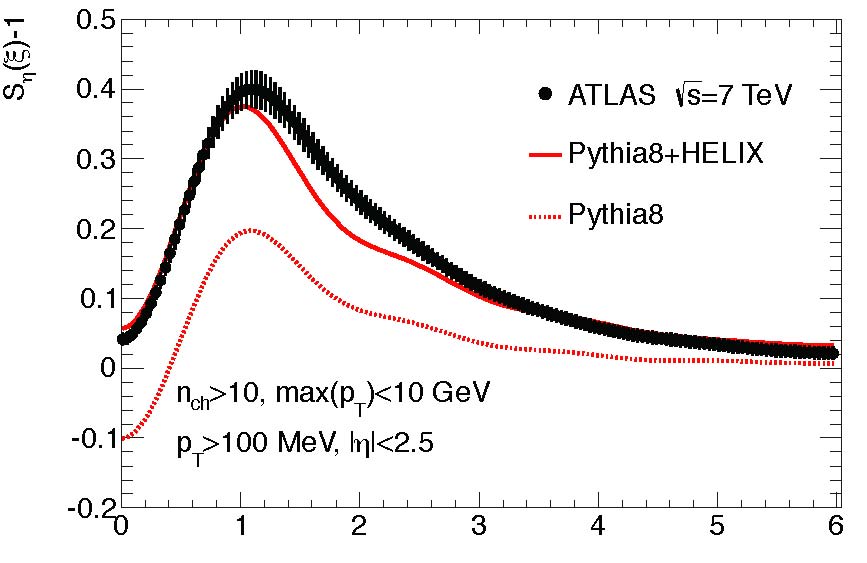}
\caption{\label{fig:helix} Helix string fragmentation describes low-p$t$ tracks better than the "standard" Lund string model.The horizontal axis - $\xi$, a dimensionless power spectrum parameter\cite{helix}.}
\end{figure}

The first indication for the helix structure in the QCD string has been found in ATLAS in an analysis of  low-$p_T$ tracks\cite{helix}. A comparison of two versions of Pythia with "standard" Lund string fragmentation and "helix" fragmentation is shown in Fig.\ref{fig:helix}. Incorporating the "helix" string structure in modeling of the non-perturbative fragmentation process should benefit all analyses. 
\subsection{\label{sec:level2}Top quark studies} 

The top quark is the heaviest particle in the SM, and it may be playing a special role in the electroweak symmetry breaking. Most  physics "beyond the SM" will appear as an excess of events above the SM predictions including the top quarks. It is imperative that the top quark production has to be understood well, as it is the most important background for many of "new physics" signatures. Top studies may also be the best testing ground for the new NNLO calculations\cite{NNLO}, which may explain the $t \bar{t}$ charge asymmetry puzzle found by CDF at the Tevatron. 

\begin{figure}[h]
\includegraphics[width=0.47\textwidth ]{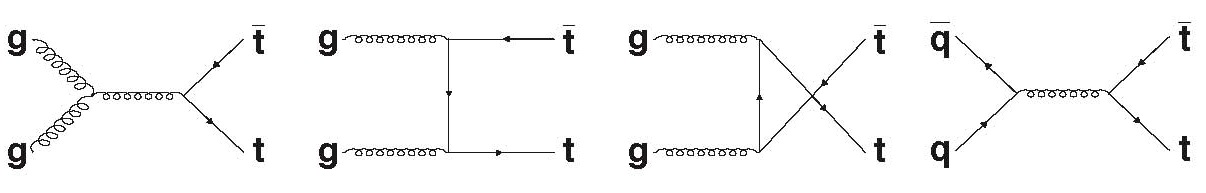}
\caption{\label{fig:FD_tt} The leading-order diagrams of top-antiquark pair production.}
\end{figure}

The leading-order diagrams  of $q\bar{q}$ and gg fusion processes leading to $t \bar{t}$  pair production are shown in Fig.\ref{fig:FD_tt}.

\begin{figure}[h]
\includegraphics[width=0.49\textwidth]{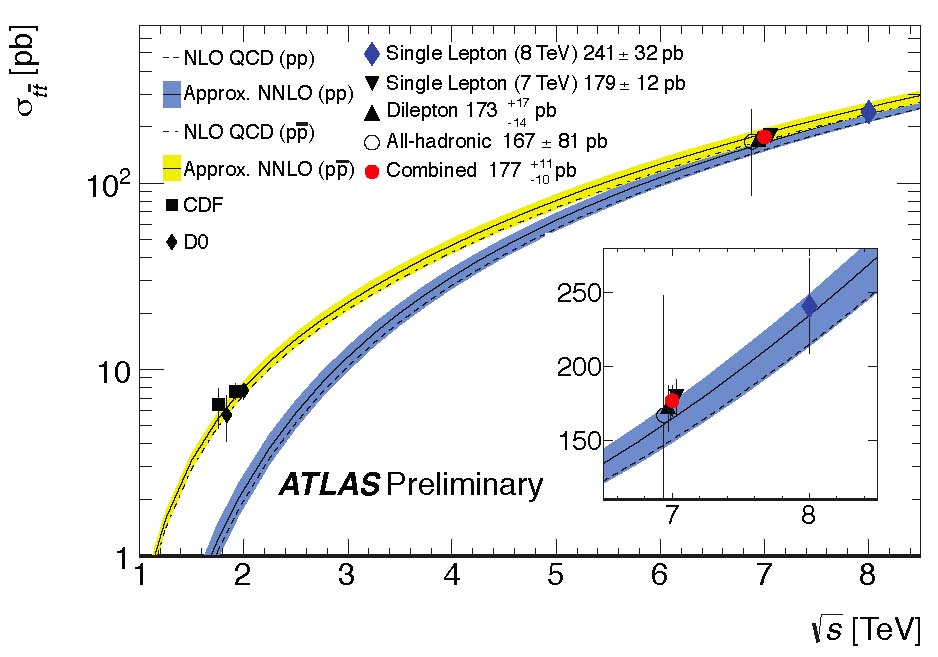}
\caption{\label{fig:TT} The top-antiquark pair production cross section measurements as a function of $\sqrt(S)$, as measured in several final states and compared with theoretical predictions\cite{top}.}
\end{figure}

The top quark, in addition to being of fundamental importance on its own, provides a unique opportunity to test QCD predictions, and to improve tuning of Monte Carlo simulations, which is essential for all analyses. ATLAS has measured the top quark production cross section in pp collisions at $\sqrt(s)=7~$TeV and $8~$TeV. The $t\bar{t}$ production cross section at has been found to be $\sigma=177^{+11}_{-10}$pb at $7$ TeV, and $\sigma=241\pm 32$pb at $8~$TeV\cite{top}. Shown in Fig.\ref{fig:TT} are the measurements  of top pair production cross sections, compared with a number of  theoretical calculations. No discrepancy between the data and the theoretical predictions is observed. 

\subsection{\label{sec:level2}ZZ studies} 

\begin{figure}[h]
\includegraphics[width=0.47\textwidth ]{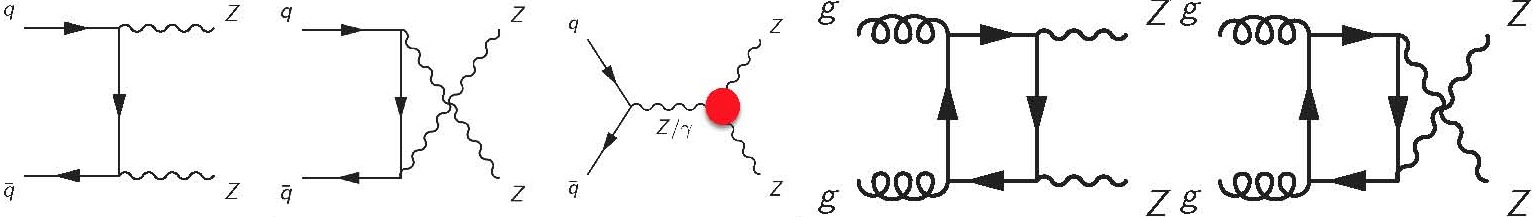}
\caption{\label{fig:FD_ZZ} The leading-order diagrams of $ZZ$ pair production.}
\end{figure}

\begin{figure}[h]
\includegraphics[width=0.49\textwidth]{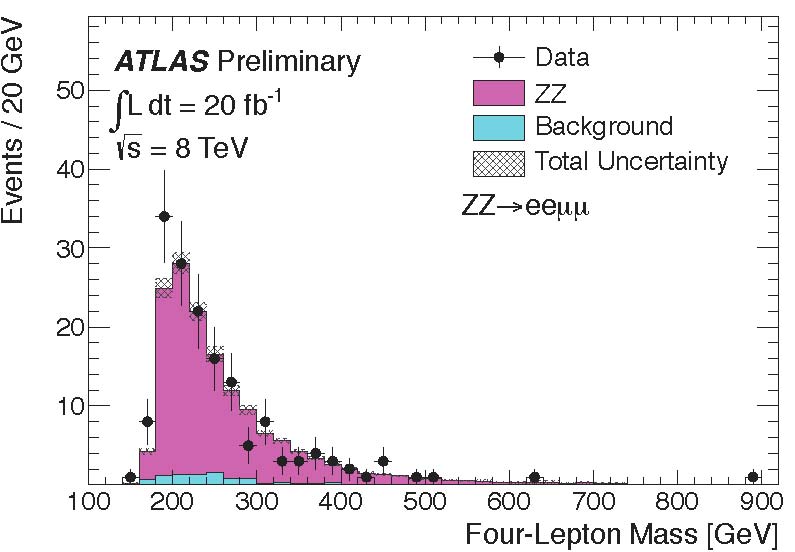}
\caption{\label{fig:ZZ_mass} Invariant mass of the four-lepton system ($\mu\mu ee$), based on 8 TeV data\cite{ZZ}.}
\end{figure}

Measurements of the electroweak bosons production and their properties provide very important tests of the SM. Deviations could indicate physics beyond the SM, for example, the neutral triple-gauge-coupling, marked with the red dot, is not allowed in SM. The leading-order diagrams of $q\bar{q}$ and gg fusion processes leading to $ZZ$  pair production are shown in Fig.\ref{fig:FD_ZZ}. The distribution of the mass of the four-lepton system is shown in Fig.\ref{fig:ZZ_mass}, compared with the background estimates. The $ZZ$ production cross section was found to be $\sigma=7.1^{+0.5}_{-0.4}(stat)\pm 0.3(syst)\pm0.2(lumi)$pb in pp collisions at $8~$TeV\cite{ZZ}. There is a good agreement between the $ZZ$ measurements and the NLO SM calculations. 

\subsection{\label{sec:level2}WZ studies} 

\begin{figure}[h]
\includegraphics[width=0.47\textwidth ]{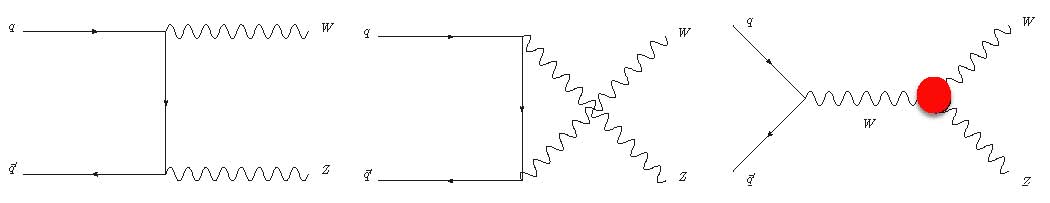}
\caption{\label{fig:FD_WZ} The leading-order diagrams of $WZ$ pair production.}
\end{figure}

\begin{figure}[h]
\includegraphics[width=0.50\textwidth]{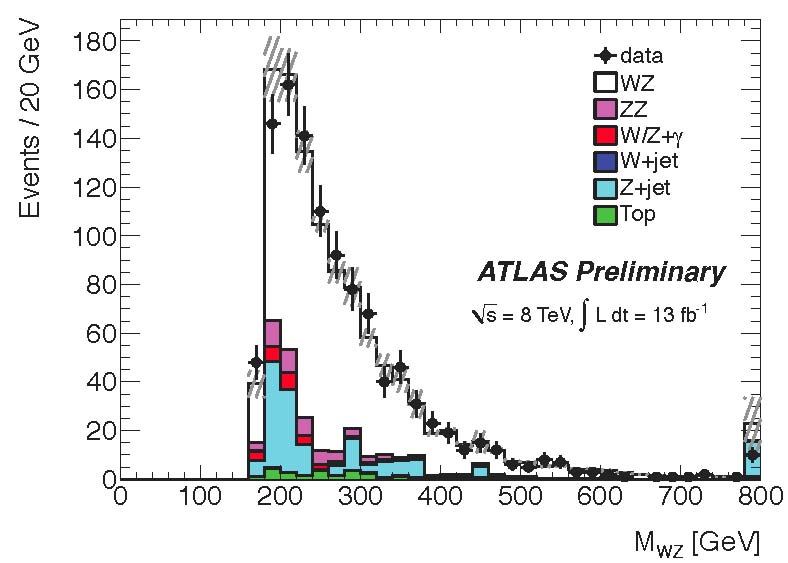}
\caption{\label{fig:WZ_mass} Invariant mass of the $WZ$ system, based on 8 TeV data\cite{WZ}.}
\end{figure}

The tree-level diagrams of processes leading to $WZ$  pair production are shown in Fig.\ref{fig:FD_WZ}, and the distribution of the mass of the $WZ$ system is shown in Fig.\ref{fig:WZ_mass}, together with the background estimates. The $WZ$ production cross section has been measured\cite{WZ} to be $\sigma=20.3^{+0.8}_{-0.7}(stat)^{+1.2}_{-1.1}(syst)^{+0.7}_{-0.6}(lumi)$pb in pp collisions at $8~$TeV. Again, there is a good agreement between the measurements and the NLO SM calculations. Deviations could indicate physics beyond the SM. In the $WZ$ case, the triple-gauge-coupling, marked with the red dot, is allowed in SM.

\begin{figure}[h]
\includegraphics[width=0.50\textwidth]{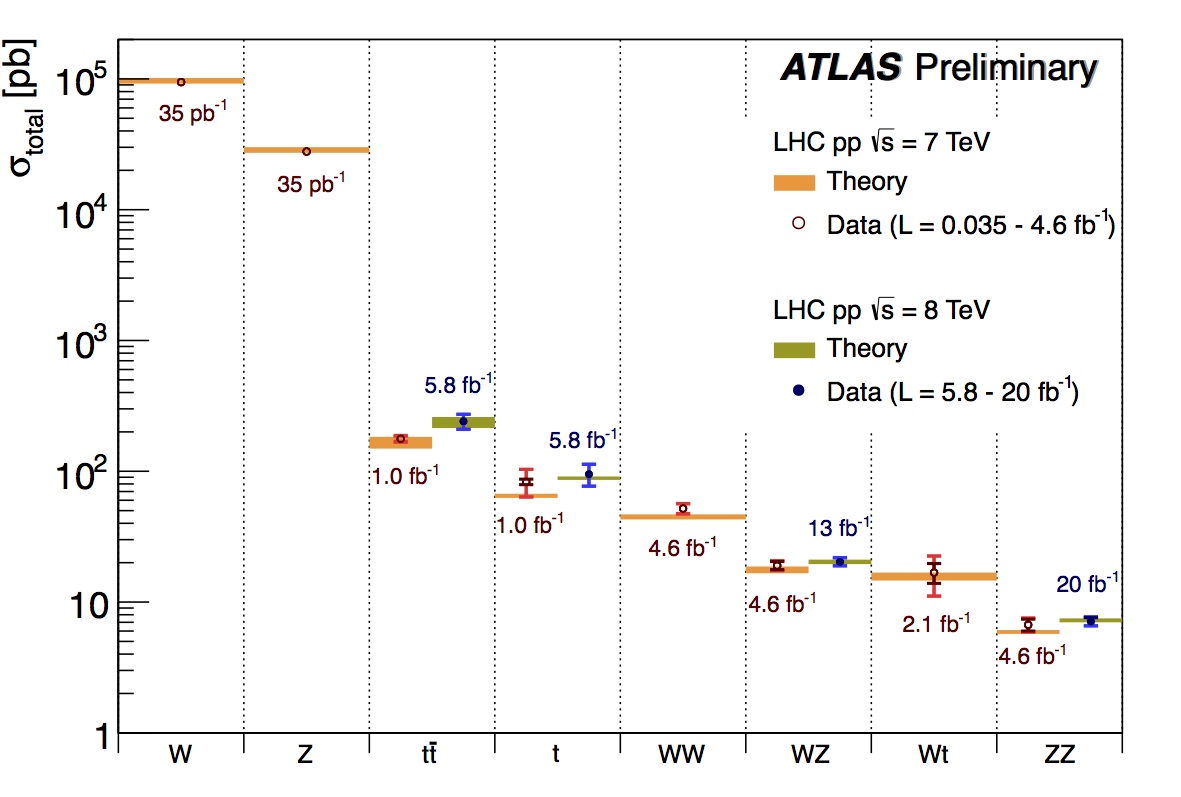}
\caption{\label{fig:EWK} Summary of Standard Model total production cross section measurements. The W and Z vector-boson inclusive cross sections were measured with 35/pb of integrated luminosity from the 2010 dataset. All other measurements were performed using the 2011 dataset or the 2012 dataset. The dark-color error bar represents the statistical uncertainty; the lighter-color error bar represents the full uncertainty, including systematics and luminosity uncertainties\cite{twiki}.}
\end{figure}

The status of the ATLAS electroweak and top production cross section measurements is presented in Fig.\ref{fig:EWK}. For details and more information about the ATLAS measurements other than $t\bar{t}, ZZ$, and $WZ$, see Marilyn Marx's talk at this conference\cite{marilyn}

\subsection{\label{sec:level2}Supersymmetry}

SUSY particles are expected to be produced strongly in pairs, and their decay chains invariably include a lightest supersymmetric particle (LSP), which is usually neutral. The typical signatures include final states with a variable number of jets, also multi-leptons, same-sign leptons - almost always with large \MET. A large mixing in the 3rd generation of SUSY fermions is expected, with at least one the top squarks expected to be light.

The Minimal Supersymmetric SM is difficult to reconcile with M$_H=$125 GeV - a smaller mass is predicted for its lightest neutral Higgs, h. Recently, it has been shown that the radiative corrections due to SUSY particles increase the h mass limits, making SUSY compatible with $M_H=125$ GeV and all other data. However, a large part of the MSSM parameter space is excluded. In more complicated models, like the NMSSM, a singlet chiral superfield added to MSSM allows the alleviation of the lightest Higgs mass (H$_1$) problem. This model has 7 physical Higgs particles (MSSM has 5).

Many searches for physics beyond the SM look for SUSY particles. I'll show only a few latest results - for more 
details and more results see Carsten Hensel's talk\cite{carsten}.
\subsubsection{\label{sec:level3}SUSY: stop searches}

\begin{figure}[h]
\includegraphics[width=0.50\textwidth]{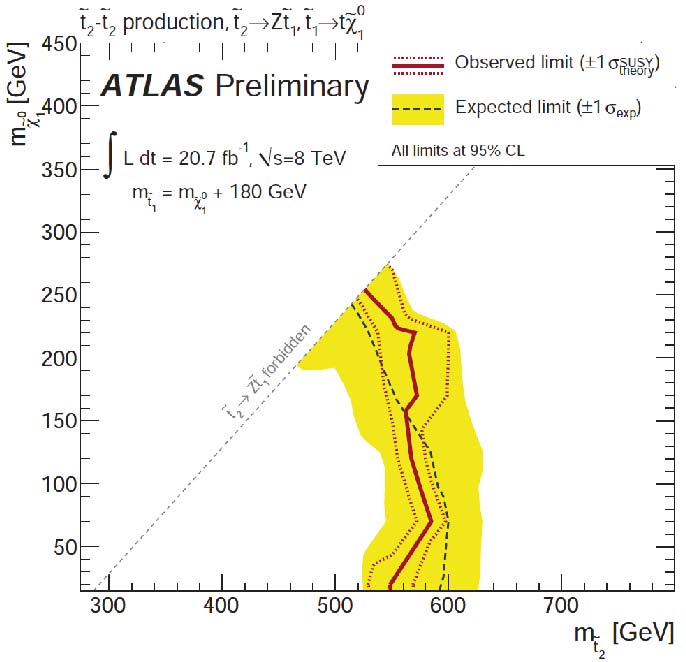}
\caption{\label{fig:susy_t2t2} Limits on $\tilde{t_2}$ stop, based on 8 TeV data\cite{A25}.}
\end{figure}

\begin{figure}[h]
\includegraphics[width=0.50\textwidth]{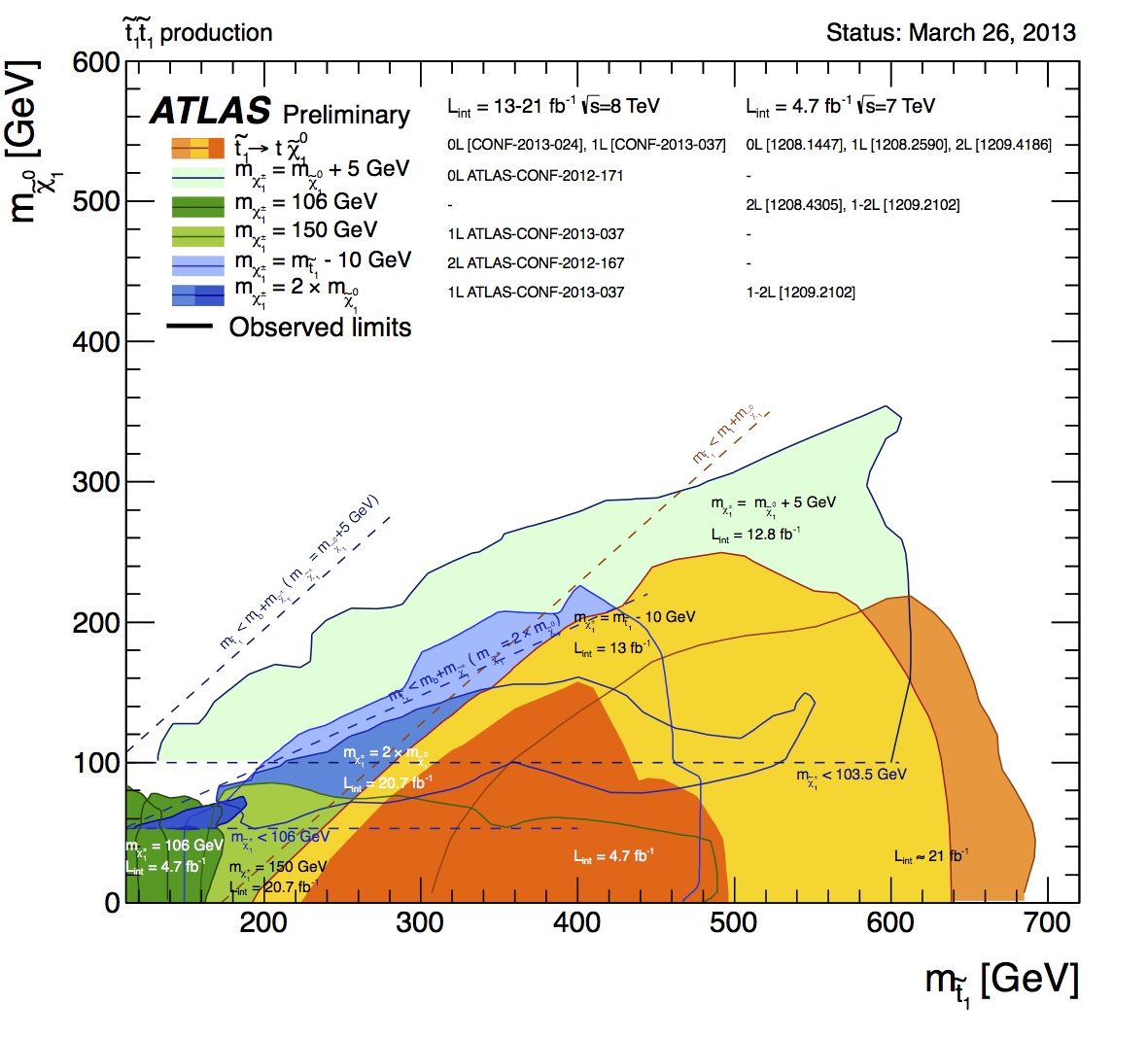}
\caption{\label{fig:susy_t1t1} These overlay contours belong to different $\tilde{t_1}$ stop decays, different sparticle mass hierarchies, and simplified decay scenarios - they should be interpreted with care\cite{twiki}. All limits are shown at 95\% C.L. }
\end{figure}

The results of an analysis\cite{A25} looking for $\tilde{t_2}$ stop pair production in which both stops decay into a $\tilde{t_1}$ stop quark and Z, $\tilde{t_2} \tilde{t_2} \to \tilde{t_1} \to Z \tilde{t_1},\tilde{t_1} \to  t \tilde{\chi^0_1}$, are shown in Fig.\ref{fig:susy_t2t2}. The exclusion results are shown in the neutralino mass - stop mass plane, (m$(\tilde{\chi^0_1}$) vs m($\tilde{t_1}$)).  Another analysis looked for stop pair production in the final states in which both stops decay into top quark and and LSP\cite{A24}, 
$\tilde{t_1} \tilde{t_1} \to t \tilde{\chi^0_1}t \tilde{\chi^0_1}$,
with both t quarks decaying hadronically - leading to multijets+\MET signature. Assuming a massless LSP, the range of 320 $GeV>m(stop)>660$ GeV has been excluded. For m(LSP)=150 GeV, 400 $GeV>m(stop)>620$ GeV, with both limits set at @95\% C.L. The third analysis looked at the decay chain 
$\tilde{t_1} \tilde{t_1} \to \tilde{t_1} \to b \tilde{\chi_1^{\pm}},\tilde{\chi_1^{\pm}} \to  t W^*{\chi^0_1}$. A summary of exclusion limits for the $\tilde{t_1} \tilde{t_1}$ searches is shown in Fig. \ref{fig:susy_t1t1}.
 

\subsubsection{\label{sec:level3}SUSY: squarks and sgluino searches} 
 
\begin{figure}[h]
\includegraphics[width=0.49\textwidth]{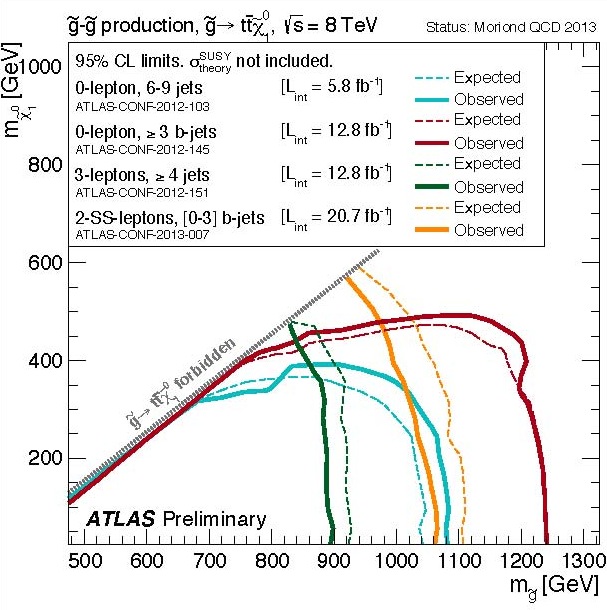}
\caption{\label{fig:susy_gg} Exclusion contours on gluino mediated stop production, based on 8 TeV data. All limits\cite{A07} are at 95\% C.L.}
\end{figure}

Squark and sgluinos are expected to be produced in pairs via strong interactions. Among the wide range of expected experimental signature for those super particles are events with zero leptons, same-sign dileptons or three leptons, all with varying number of jets and tagged b-jets.  A summary of searches\cite{A07} looking for gluino mediated stop production are shown in Fig.\ref{fig:susy_gg}.

\begin{figure}[h]
\includegraphics[width=0.50\textwidth]{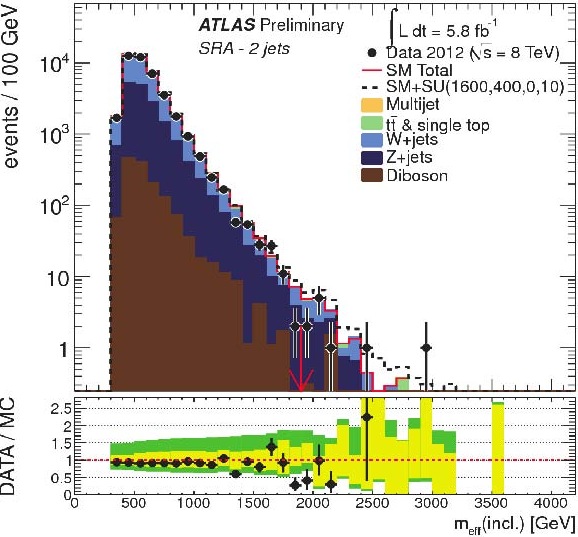}
\caption{\label{fig:susy_2j6j} Inclusive mass distributions of the multijet + \MET system\cite{109}.}
\end{figure}

Searches for direct production and decays of the 1st and 2nd generation of squarks and gluinos\cite{109} are based on events with 2 jets + \MET, 4 jets + \MET and 6 jets + \MET in the final state. 
No excess of events has been found, and model dependent limits on the masses of sparticles, O($1~$TeV), have been set.
Fig.\ref{fig:susy_2j6j} presents results of such a search in events with 2 jets + \MET.

\subsection{\label{sec:level2} Exotic searches} 
 
\begin{figure}[h]
\vspace{0.3cm}
\includegraphics[width=0.5\textwidth]{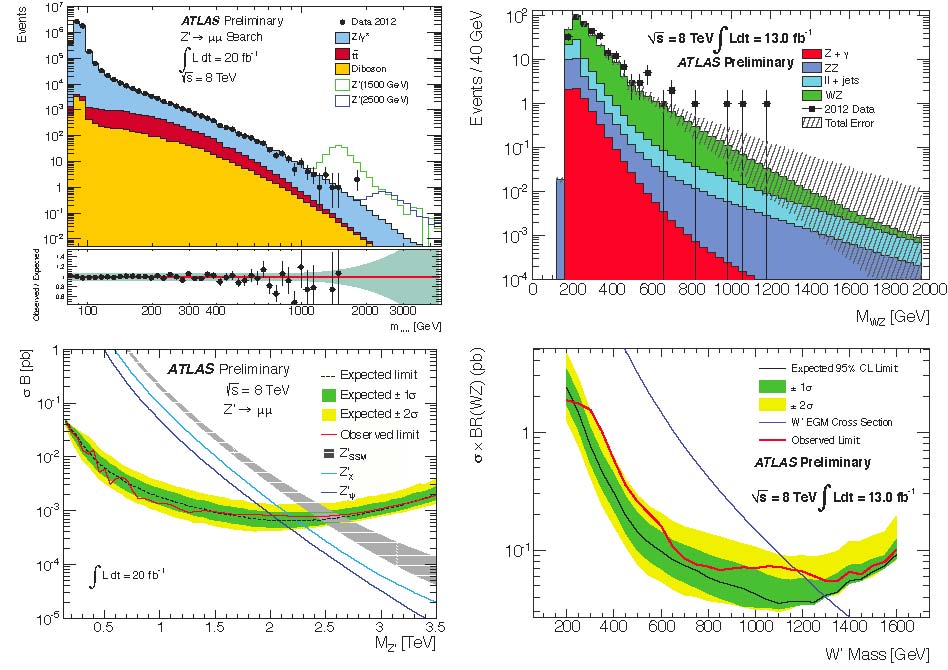}
\caption{\label{fig:exotics_z_w} Searches for $Z'$ and $W'$ production in ATLAS 8 TeV data\cite{A17}\cite{A15}.}
\end{figure}

Many searches for the final states not expected in the SM, observation of which could point to new physics, fall in this category. The simplest analyses look for resonances - peaks in the invariant mass distributions of dileptons ($Z' \to \mu^+\mu^-, e^+e^-$), dibosons, three leptons and \MET (W'). Enhanced production of prompt photons ($\gamma\gamma$) or like-sign leptons (for example - $\mu^{\pm}\mu^{\pm}$) would constitute one of the most powerful of the possible indications of new physics.
Example of such searches for the new $Z'$\cite{A17} and $W'$\cite{A15} bosons are shown in Fig.\ref{fig:exotics_z_w}. As a result of combining the results of the searches for  $Z' \to e^+e^-$ and  $Z' \to \mu^+\mu^-$, a limit for M($Z')> 2.86 $ TeV  @95\% C.L. has been set. These results can be interpreted in many other, model-dependent, ways.

\begin{figure*}
\includegraphics{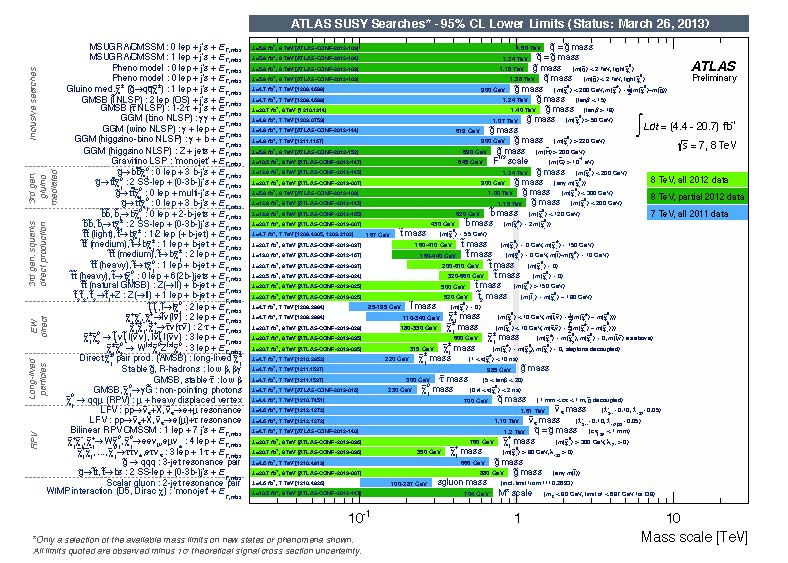}
\caption{\label{fig:susy_all} A summary of mass reach of ATLAS searches for Supersymmetry. Only a representative selection of the available results is shown\cite{twiki}. }
\end{figure*}

\begin{figure*}
\includegraphics{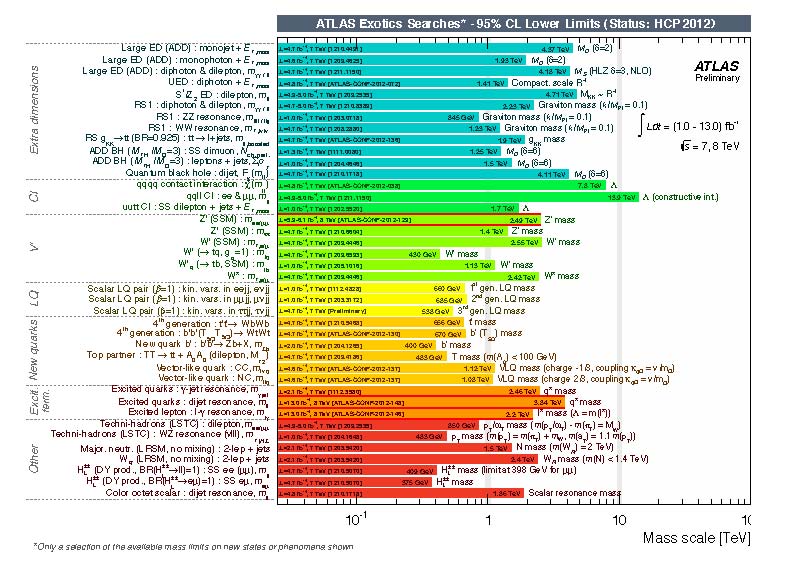}
\caption{\label{fig:exotics_all} Mass reach of ATLAS searches for new phenomena other than Supersymmetry. Only a representative selection of the available results is shown\cite{twiki}.}
\end{figure*}

A  summary of the mass limits obtained in the ATLAS SUSY searches is presented in Fig.\ref{fig:susy_all}, and a corresponding list of  limits for the Exotic searches is shown in Fig.\ref{fig:exotics_all}.

\subsection{\label{sec:level2} MSM Higgs searches and Higgs properties} 

\begin{figure}[h]
\includegraphics[width=0.49\textwidth]{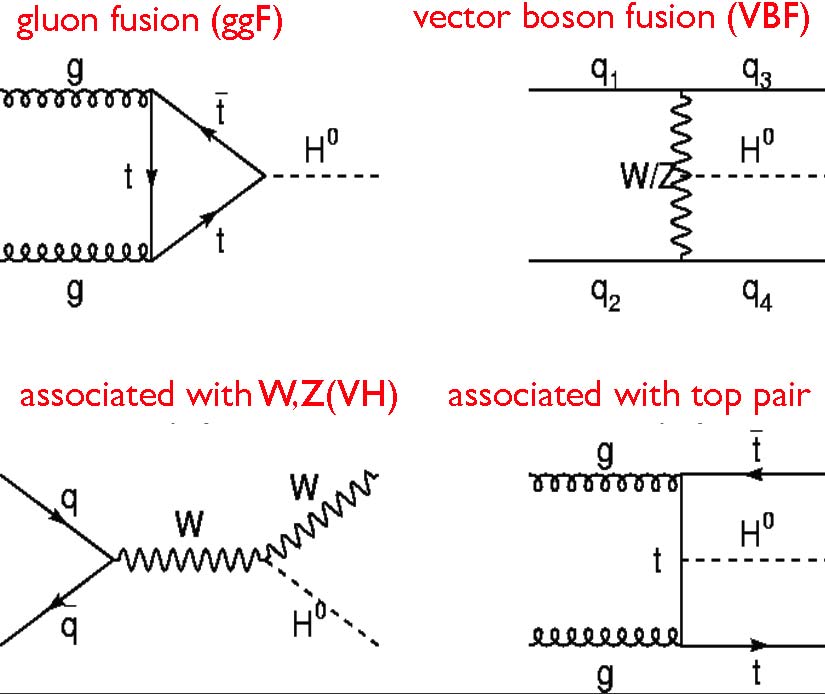}
\caption{\label{fig:FD_higgs} The four dominant Higgs production processes in the MSM. }
\end{figure}

\begin{figure}[h]
\includegraphics[width=0.49\textwidth]{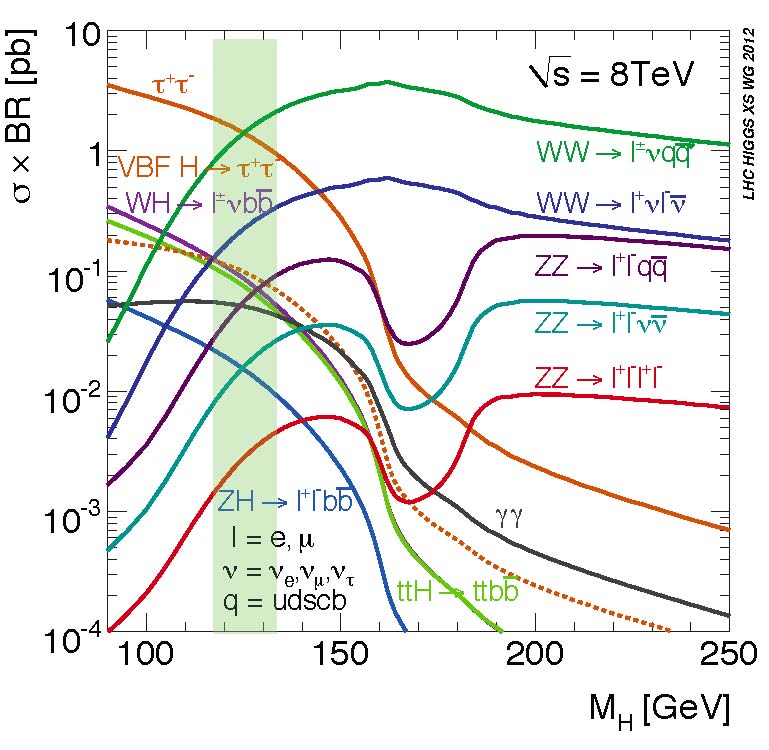}
\caption{\label{fig:higgs_sigma} The product of the Higgs cross section and its branching fractions to different final states, according to the MSM. }
\end{figure}

\begin{figure}[h]
\includegraphics[width=0.49\textwidth]{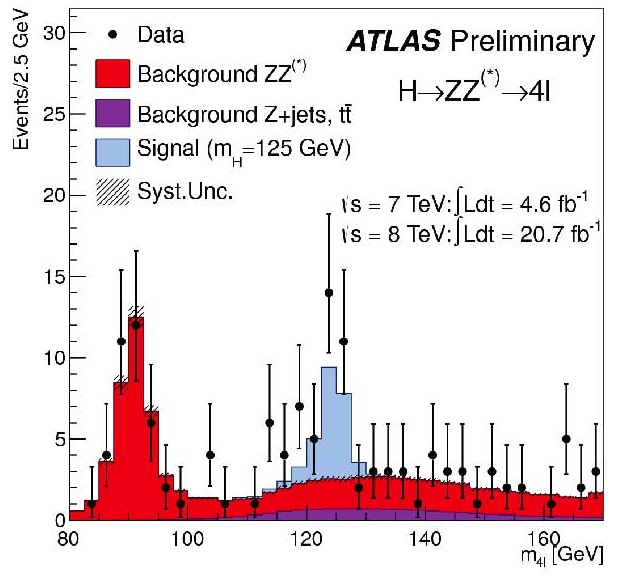}
\caption{\label{fig:higgs_mass_4l} The mass distributions of of the 4 leptons system in ATLAS events - Higgs boson candidates\cite{A13}.}
\end{figure}

\begin{figure}[hf]
\includegraphics[width=0.49\textwidth]{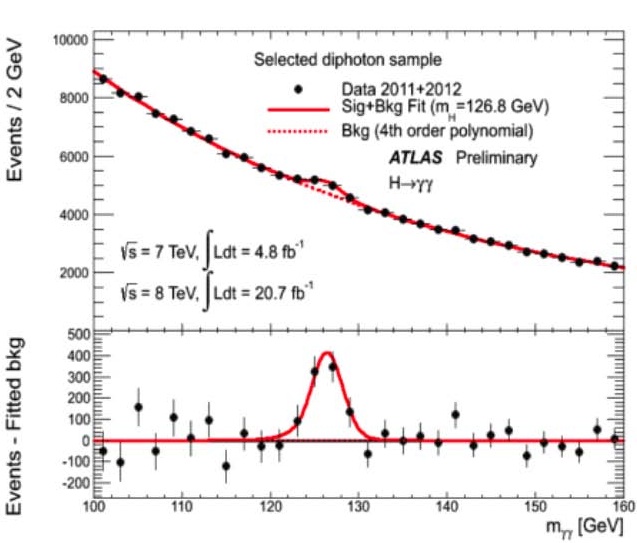}
\caption{\label{fig:higgs_mass_gg} The mass distributions of of the $\gamma\gamma$ system in ATLAS events - Higgs boson candidates\cite{A12}.}
\end{figure}

\begin{figure}[hf]
\includegraphics[width=0.50\textwidth]{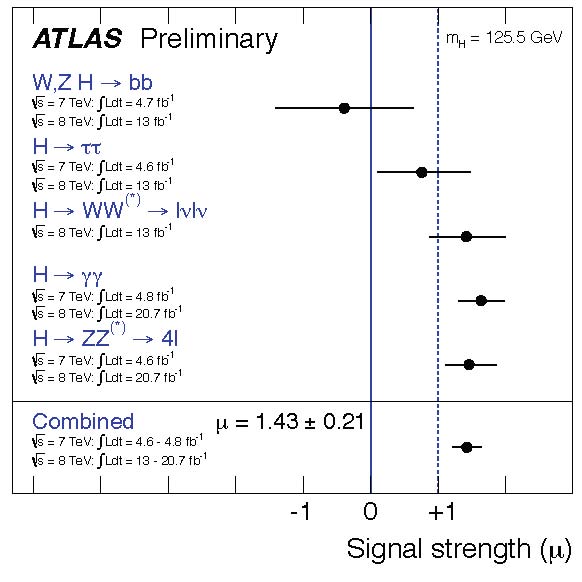}
\caption{\label{fig:higgs_mu} Signal strength as a function of the Higgs mass for $H\to \gamma\gamma$, $H \to ZZ^*$ and their combination\cite{A14}. }
\end{figure}

The most important Higgs boson production processes in the MSM are: 
\begin{itemize}
\item gluon fusion (ggF) - has the largest cross sections, but also large backgrounds
\item vector boson fusion (VBF) - has smaller cross section but also smaller backgrounds
\item associated production with a pair of $W,Z$ bosons (VH) - has smaller cross section but also smaller backgrounds
\item associated production with a top quark pair
\end{itemize}
The lowest-order Feynman diagrams of these four production processes are shown in Fig.\ref{fig:FD_higgs}.
At pp collisions at 8 TeV, the Higgs production cross section is about 30\% higher than at 7 TeV.  However, the increase of sensitivity resulting from an increase in the available energy in the pp collisions from 7 TeV to 8 TeV is less, only a factor of 1.1-1.15, as the irreducible backgrounds increase a bit less than the Higgs production, but the reducible backgrounds (top and Zbb) increase by a bit more. After taking into account the branching fractions of the various Higgs decay modes, the most sensitive channels (for 120 GeV$<M_H<130$ GeV) are: 
\begin{itemize}
\item $H \to ZZ^* \to 4$ leptons
\item $H \to \gamma\gamma$
\item $H \to WW^* \to l\nu l\nu$
\item $H \to \tau \tau$
\item $W/Z+H \to W/Z+bb$
\end{itemize}
By some strange coincidence, $M_H=125$ GeV is one of the best places, from the experimental point of view, to find the Higgs and to study its properties. A large number of channels with relatively large branching fractions are available in which to observe the Higgs boson and to compare their measured production rates with the MSM predictions. Fig.\ref{fig:higgs_sigma} presents the corresponding fractional cross sections in different final states, $\sigma \times$ BF. Both the $\gamma\gamma$ and the four-lepton final states allow the measurement of the Higgs mass with very high precision. The corresponding mass plots are shown in Fig.\ref{fig:higgs_mass_4l} and Fig.\ref{fig:higgs_mass_gg} - both show peaks with significance $>5.9\sigma$. In the four-lepton final state\cite{A13}, the peak position is at $M_H=124.3^{+0.6}_{-0.5}(stat)^{+0.5}_{-0.3}(syst)$ GeV, while in the $\gamma\gamma$ channel\cite{A12}  the peak in the mass distribution is at $M_H=126.8\pm 0.2 (stat) \pm 0.7 (syst)$ GeV. The combined Higgs mass value is $M_H=125.5\pm 0.2(stat)^{+0.5}_{-0.6}(syst)$ GeV. The two values are not the same, they differ by $\Delta M_H=2.3^{+06}_{-0.7}(stat)\pm0.6(syst)$ GeV. The hypothesis that the mass values obtained in the two channels are identical, $\Delta M_H=0$, is not favored at the level of 2.5$\sigma$ (1.2\%) to 8\%, depending on the assumptions in the treatment of errors and on the statistical analysis itself. Assuming that the new boson is the MSM Higgs, all the couplings and Higgs branching fractions can be calculated in the model. Thus, the relative abundance of the Higgs signal in different final states, usually measured as the ratio of the observed rate of events to the predicted rate of events, $<\mu>$, is of fundamental importance. The signal strengths, $<\mu>$ for a number of final states\cite{A14} are shown if Table \ref{tab:table1}, and in Fig.\ref{fig:higgs_mu} the signal strengths for the two most sensitive channels, $H \to \gamma\gamma$ and $H\to ZZ^*$, are shown as a function of Higgs mass. ATLAS observes a bit more events that expected, however, the probability of the average $<\mu>$ to be compatible with the MSM is an acceptable 9\%. Both discrepancies, the differing mass values in the $\gamma\gamma$ and $ZZ^*$ final states, and the larger than expected value of $<\mu>$, are intriguing, and they will certainly be carefully re-examined with more data in the next LHC run which will begin in 2015. 

\begin{figure}[hf]
\includegraphics[width=0.50\textwidth]{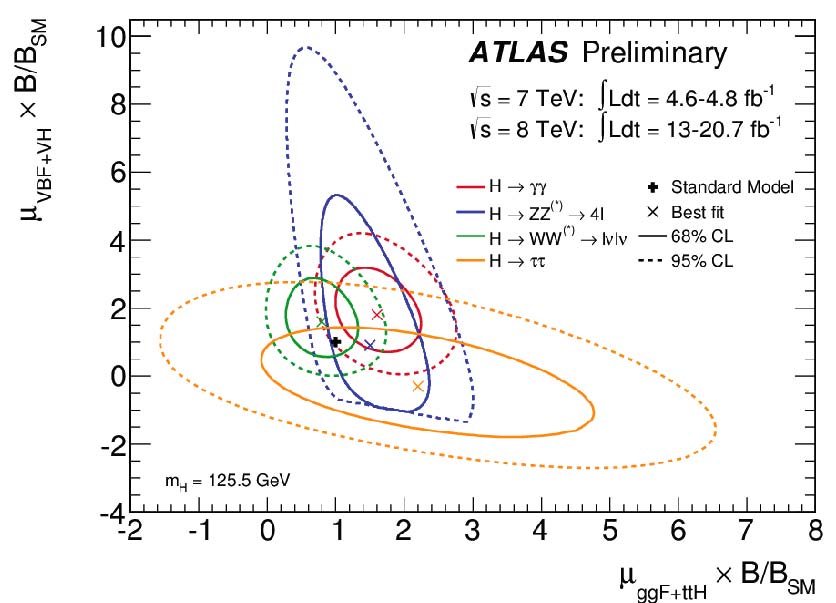}
\caption{\label{fig:higgs_vbf}The contours in the signal strengths $<\mu>_{VBF+VH}$ vs $<\mu>_{ggF+ttH}$ plane, assuming MSM and $M_H=125.5$ GeV\cite{A14}}
\end{figure}

The question of whether the new boson is the MSM Higgs can also be examined by comparing the relative strength of the fermion and boson couplings. Measurements of the relative production rates is crucial for establishing properties of the new boson. Assuming the MSM Higgs, the expected cross sections for $M_H=125$ GeV at 8 TeV are:
\begin{itemize}
\item ggF - 19.5 pb, fermion couplings (accessible in $\gamma\gamma$, $ZZ^*$ and $WW^*$ decay final states ) 
\item VBF - 1.6 pb, boson couplings ( $\gamma\gamma$, $ZZ^*$ and $WW^*$ with 2 or more jets) - 3.1$\sigma$ evidence in ATLAS data
\item VH  - 1.1 pb, boson couplings ( $\gamma\gamma$, $ZZ^*$ and $WW^*+W,Z$ 
\item ttH  - 0.1 pb, fermion couplings
\end{itemize}
The contours in the signal strengths $<\mu>_{B}$ vs $<\mu>_{F}$ plane are shown Fig.\ref{fig:higgs_vbf}. 

\begin{figure}[hf]
\includegraphics[width=0.50\textwidth]{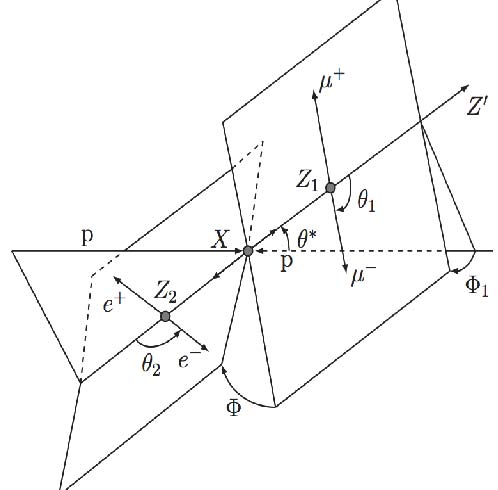}
\caption{\label{fig:higgs_spin}The definition of angles used in a measurement of the Higgs spin in the $H \to ZZ^* \to$ 4 leptons final state\cite{A13}. }
\end{figure}

\begin{table}
\caption{\label{tab:table1}Signal strength $<\mu>$ for different Higgs decay modes, assuming the MSM Higgs with $M_H=125$ GeV.}
\begin{ruledtabular}
\begin{tabular}{cc}
 Higgs boson decay&$<\mu>$ \\  \hline
 $VH \to Vbb$&$-0.4\pm 1.0$  \\
 $H \to \tau\tau$&$0.8\pm 0.7$  \\
 $H \to WW^*$&$1.0\pm 0.3$  \\
 $H \to \gamma\gamma$&$1.6\pm 0.3$  \\
 $H \to WW^*$&$1.5\pm 0.4$  \\ \hline
 Combined&$1.30\pm 0.20$  \\ 
\end{tabular}
\end{ruledtabular}
\end{table}
In the MSM, the spin-parity of the Higgs boson is expected to be $J^P=0^+$. The spin information has been extracted from ATLAS data by analyzing the angular distribution of the Higgs decay products in three final states: $\gamma\gamma$, $ZZ^* \to 4l$ and $WW^*$. As an illustration of the procedure, the definition of the relevant angles in the $H\to ZZ^* \to$ 4 leptons decay is shown in Fig.\ref{fig:higgs_spin}. The ATLAS results are:
\begin{itemize}
\item $H\to \gamma\gamma$ - spin 2 excluded at 2.8 $\sigma$ - assuming 100\% gg, spin 1 is obviously excluded, as well
\item$H \to$ 4 leptons - spin $0^-$ excluded at $>2\sigma$, spin 2 excluded at 1.5-3$\sigma$ - assuming 0-100\% gg
\item$H \to WW^*$ - spin 2 excluded at 95-99 \% C.L. - depending on the assumed fraction of gg
\end{itemize}
Although it is too early to make definite statements, the ATLAS measurements of the spin-parity of the new boson strongly suggest that the new particle looks like a MSM Higgs. More details about the ATLAS Higgs analyses can be found in Sofia Maria Consonni's talk\cite{sofia}.
\section{\label{sec:level1} Future plans} 

According to the current schedule of the LHC accelerator, the next physics run will start in 2015, after Phase-0 upgrade. The LHC will reach the pp energy of $\sim$ 13 TeV, and the luminosity of $1\times10^{34}s^{-1}cm^{-2}$ with the bunch spacing of 25 ns. The expected integrated luminosity in this second run is 75-100/fb. The physics run will be followed by another long shutdown, in which the LHC (and the detectors) will undergo the Phase-1 upgrade to the full energy (14 TeV) and the design luminosity - $2\times10^{34}s^{-1}cm^{-2}$ with 25 ns bunch spacing. In the  physics run which will follow, the experiments will record about 350/fb of data. It has been estimated that with $\sim$300/fb of data after Phase-1 upgrade, the spin-parity of Higgs will be known to $\sim 5 \sigma$ level, and the ratios of couplings will be known to within 30-50\%. To determine the fine details of the Higgs potential, the multi-Higgs couplings have to be measured. The $HHH$ couplings may be reachable with the data corresponding to the integrated luminosity of 3000/fb, after the Phase-2 upgrade to HL-LHC - perhaps only in 2030. However, the $HHHH$ couplings will be, most likely, beyond the reach of the LHC.

Of course, with the energy increase from 8 to 13 TeV, in addition to Higgs boson(s) studies, there will be another round of comprehensive searches for SUSY and other "new physics". This is what the physics goal of the LHC program is - to explore the new, previously unattainable, energies, and - in turn - new regions of phase space and model parameter space. 

Finding the new boson was a {\it great} physics result. However, if it is just the Minimal Standard Model Higgs boson - the simplest possible realization of the electroweak symmetry breaking - it will leave many unanswered questions, and the fine tuning problem (or gauge hierarchy problem) will still be with us. 
It is possible that with an increase of the pp collision energy from 8 TeV to $13$ TeV we will cross a threshold above which we will observe new particles, too heavy to have been observed so far. This would be  {\it really great}.

If not, then perhaps we will have to turn our attention to precise measurements of the branching fractions and properties of the Higgs boson, either at the LHC, or at a new $e^+e^-$ collider, a {\it cleaner} environment in which to study the MSM Higgs boson.

\newpage



\end{document}